\documentclass[useAMS,usenatbib]{mn2e}
\usepackage{graphicx} 
\usepackage{graphicx}
\usepackage{longtable}
\usepackage{natbib}
\usepackage{lscape}
\usepackage{color}

\newcommand{\mnras}{MNRAS\,\,}

\newcommand{\aap}{A\&A\,\,}

\newcommand{\apj}{ApJ\,\,}

\newcommand{\aj}{AJ\,\,}

\newcommand{\apss}{Ap\& SS\,\,}

\newcommand{\actaa}{Acta Astron.\,\,}

\title[]{The VMC Survey -  XX. Identification of new Cepheids in the Small Magellanic 
Cloud\thanks{Based on observations collected at the European Organisation for Astronomical Research in the Southern Hemisphere under ESO programme(s) 179.B-2003.}}
\author[M. I. Moretti et al.]{M. I. Moretti$^{1,2,3}$\thanks{E-mail:
mariaidamoretti@noa.gr}, G. Clementini$^{1}$, V. Ripepi$^{4}$, M. Marconi$^{4}$, S. Rubele$^{5,6}$,
 \newauthor  M.-R.L. Cioni$^{7,8,9}$, T. Muraveva$^{1}$, M. A. T. Groenewegen$^{10}$, N. J. G. Cross$^{11}$
 \newauthor  V.D. Ivanov$^{12,13}$, A. E. Piatti$^{14,15}$, R. de Grijs$^{16,17,18}$  
\\
$^{1}$ INAF-Osservatorio Astronomico di Bologna, via Ranzani 1, 40127 Bologna, Italy \\ 
$^{2}$ Scuola Normale Superiore di Pisa, piazza dei Cavalieri 7, 56126 Pisa, Italy \\
$^{3}$ IAASARS, National Observatory of Athens, 15236 Penteli, Greece\\
$^{4}$ INAF-Osservatorio Astronomico di Capodimonte, via Moiariello 16, 80131 Napoli, Italy \\
$^{5}$ INAF-Osservatorio Astronomico di Padova, vicolo dell'Osservatorio 5, 35122 Padova, Italy \\ 
$^{6}$ Dipartimento di Fisica e Astronomia, Universit\`a di Padova, Vicolo dell'Osservatorio 2, I-35122 Padova, Italy \\ 
$^{7}$ Universit\"{a}t Potsdam, Institut f\"{u}r Physik und Astronomie, Karl-Liebknecht-Str. 24/25, 14476 Potsdam, Germany \\
$^{8}$ Leibnitz-Institut f\"{u}r Astrophysik Potsdam, An der Sternwarte 16, 14482 Potsdam Germany \\ 
$^{9}$ University of Hertfordshire, Physics Astronomy and Mathematics, College Lane, Hatfield AL10 9AB, United Kingdom \\
$^{10}$ Royal Observatory of Belgium, Ringlaan 3, B-1180 Brussels, Belgium\\
$^{11}$ Wide-Field Astronomy Unit, Institute for Astronomy, School of Physics and Astronomy, University of Edinburgh,
Royal Observatory,\\ Blackford Hill, Edinburgh EH9 3HJ, UK\\
$^{12}$ European Southern Observatory, Av. Alonso de Cordoba 3107, Casilla 19, Santiago, Chile\\
$^{13}$ European Southern Observatory, Karl-Schwarzschild-Str. 2, 85748 Garching bei M\"{u}nchen, Germany \\
$^{14}$ Observatorio Astron\'omico, Universidad Nacional de C\'ordoba, Laprida 854, 5000, C\'ordoba, Argentina\\
$^{15}$ Consejo Nacional de Investigaciones Cient\'ificas y T\'ecnicas, Av. Rivadavia 1917, C1033AAJ, Buenos Aires, Argentina\\
$^{16}$ Kavli Institute for Astronomy and Astrophysics, Peking University, Yi He Yuan Lu 5, Hai Dian District, Beijing 100871, China\\
$^{17}$ Department of Astronomy, Peking University, Yi He Yuan Lu 5, Hai Dian District, Beijing 100871, China\\
$^{18}$ International Space Science Institute-Beijing, 1 Nanertiao, Zhongguancun, Hai Dian District, Beijing 100190, China 
}
\begin{document}

%\date{Received x Xxx 201x / Accepted x Xxx xxx}
\date{}

%\pagerange{\pageref{firstpage}--\pageref{lastpage}} \pubyear{2002}

\maketitle

\label{firstpage}

\begin{abstract}
We present $K_\mathrm{s}$-band light curves for 299 Cepheids
in the Small Magellanic Cloud (SMC) of which 288 are new discoveries that we have identified 
using multi-epoch 
near-infrared photometry obtained by the VISTA survey of the Magellanic Clouds system (VMC). 
The new Cepheids have periods in the range from 0.34 to 9.1 days and cover the magnitude interval  
12.9 $\leq \langle K_\mathrm{s} \rangle \leq$ 17.6 mag.
 Our method was developed using variable stars previously identified by the optical microlensing 
survey OGLE. We focus on searching new Cepheids in external regions of the SMC for which
 complete VMC $K_\mathrm{s}$-band observations  are  available and no comprehensive
identification of different types of variable stars from other surveys exists yet. 

\end{abstract}


\section{Introduction}

The Visual and Infrared survey Telescope for Astronomy (VISTA, \citealt{Eme06})
  survey of the Magellanic Clouds system (VMC, \citealt{Cio11}) 
is collecting $Y, J, K_\mathrm{s}$ photometry\footnote{The  VMC system is similar to the Vega magnitude system (see, e.g., \citealt{Rub12}).}
 down to $Y$=21.9 mag, $J$=22.0 mag, $K_\mathrm{s}$=21.5 mag 
(5 $\sigma$ level on the stacked images) 
for sources in the Large Magellanic Cloud (LMC), 
the Small Magellanic Cloud (SMC), the Bridge connecting them, and a small part of the Magellanic Stream. 
The VMC images are processed by the Cambridge Astronomical Survey Unit 
(CASU\footnote{http://casu.ast.cam.ac.uk/}; \citealt{Lew10}) through the VISTA Data Flow System
(VDFS) pipeline that performs aperture photometry of the images. 
The reduced data are then further processed by the Wide Field
Astronomy Unit (WFAU)\footnote{http://horus.roe.ac.uk/vsa} in Edinburgh where the single epochs are
stacked and catalogued in the VISTA Science Archive (VSA; \citealt{Cro12}).
 As of 2016 January, the VMC survey observations are 74\% complete, with specific completion levels
 of 62\%, 95\%, 95\% and 100\% 
for the LMC, the SMC, the Bridge and the Stream, respectively.
A main aim of the VMC survey is to study the structure of the whole Magellanic system using different distance indicators. 
Among them, most notably,  are primary standard candles such as  the Cepheids and the RR Lyrae stars,  for which VMC is obtaining 
$K_\mathrm{s}$-band light curves with  13  (or more) individual epochs, with most epochs reaching a depth of $K_\mathrm{s}$=19.3 mag (5 $\sigma$ level). 
The significantly reduced amplitude of these pulsating variables in the near-infrared, with respect to the optical bands, along with the multi-epoch cadence 
of the VMC $ K_\mathrm{s}$-band observations,  allows us 
to infer accurate mean $K_\mathrm{s}$ magnitudes. Furthermore,  the near-infrared period-luminosity ($PL$) relations are intrinsically much narrower than the 
corresponding optical relations and less affected by systematic uncertainties in reddening and metal content \citep{Cap00}.
For these reasons the VMC data are very well suited to construct $PL$ relations with a high level of  precision 
and accuracy (e.g. \citealt{Rip12a,Rip12b,Rip14,Rip15};
\citealt{Mur15}).  Conversely, the smaller amplitudes compared to optical bands make it more 
difficult to identify these variables from  near-infrared data alone.
In VMC, most of the information (identification, period, variability type, etc.) needed to fold the
 $Y, J, K_\mathrm{s}$-band light curves and derive average $Y, J, K_\mathrm{s}$ magnitudes 
for the variable stars are taken from the catalogues of Magellanic Cloud variables produced by  
large microlensing optical surveys such as MACHO \citep{Alc00}, EROS \citep{Tis07} and OGLE \citep{Sos08a} that were conducted 
in the last two  decades to search for baryonic dark matter in the Milky Way. 
As described in \cite{Uda15}, the observations of the fourth phase of the OGLE project (hereinafter,
 OGLE~IV) have been successfully run over the last five years. First results have been published in \cite{Sos12} (South Ecliptic Pole region), 
 \cite{Koz13} (Magellanic Bridge) and \cite{Sos15a} (anomalous Cepheids).\footnote{During the revision phase of this manuscript,
results on classical Cepheids in the Magellanic Clouds have been published by \cite{Sos15b}. See Section~\ref{sec:using}.}

\begin{table*}
\caption{Information on the $K_\mathrm{s}$-band photometry  of the VMC tiles analyzed in this paper.
Col. 1: Field and tile number, Col. 2: Right Ascension (RA, J2000), Col. 3: Declination (Dec, J2000), Col. 4:  position
angle, Col. 5: number of $K_\mathrm{s}$ observations, Cols. 6,7: Dates of the 1st and
last observations expressed in dd/mm/yy format, Col. 8: Time interval (T.I.) between first and last observations, 
Col. 9: Airmass, Col. 10: Full Width at Half Maximum (FWHM),
Col. 11: Ellipticity, Col. 12: Limiting magnitude.
Reported values of airmass, FWHM, ellipticity and limiting magnitude are averages over all nights. 
They are indicated with their respective standard deviations. 
The number of epochs ($N_{K_\mathrm{s}}$) includes 
both shallow and deep observations (see text for details).}
\scriptsize
\label{tab:qc}
\[
\begin{array}{cccccccrcccc}
\hline \hline
\noalign{\smallskip}
\mathrm{Tile} & \alpha & \delta & \mathrm{\phi} & N_{K_\mathrm{s}} & \mathrm{1st\,Epoch} & 
\mathrm{Last\,Epoch} & \mathrm{T.\,I.} & \mathrm{Airmass} & \mathrm{FWHM}
& e & \mathrm{LimMag}\\
& \mathrm{(h:m:s)} & (^\circ \mathrm{:}^\prime\mathrm{:}^{\prime\prime}) &
\mathrm{(deg)} &  & & & \mathrm{(d)} & & & & \mathrm{(mag)}\\
\noalign{\smallskip}
\hline
\noalign{\smallskip}

\mathrm{LMC\,6\_8} & 06\mathrm{:}02\mathrm{:}21.984 &
-69\mathrm{:}14\mathrm{:}42.360 & -83.7904 & 15 & 23/02/11 & 16/04/14 & 1148 &
1.49\pm0.04 & 0.96\pm0.19 & 0.06\pm0.01 & 19.30\pm0.24\\
\noalign{\smallskip}

\mathrm{LMC\,7\_3} & 05\mathrm{:}02\mathrm{:}55.200 &
-67\mathrm{:}42\mathrm{:}14.760 & -97.7044 & 16 & 12/01/11 & 04/03/13 & 782 &
1.44\pm0.05 & 0.95\pm0.10 & 0.06\pm 0.02 & 19.23\pm0.19\\
\noalign{\smallskip}

\mathrm{LMC\,8\_8} & 05\mathrm{:}59\mathrm{:}23.136 &
-66\mathrm{:}20\mathrm{:}28.680 & -84.4802 & 16 & 14/11/09 & 26/11/10 & 377 &
1.39\pm0.05 & 0.93\pm0.11 & 0.06\pm0.02 & 19.36\pm0.27\\
\noalign{\smallskip}

                                    &                                                               &          &     &     &     &    &   &  & & &\\

\mathrm{SMC\,3\_1} & 00\mathrm{:}02\mathrm{:}39.912 &
-73\mathrm{:}53\mathrm{:}31.920 & -11.3123 & 16 & 21/08/12 & 01/07/14 & 679 &
1.60\pm 0.05 & 1.01\pm 0.15 & 0.05\pm 0.01 & 19.38\pm0.16\\
\noalign{\smallskip}

\mathrm{SMC\,3\_3} & 00\mathrm{:}44\mathrm{:}55.896 &
-74\mathrm{:}12\mathrm{:}42.120 & -1.2120 & 18 & 03/08/11 & 05/09/12 & 399 &
1.62\pm 0.04 & 1.02\pm 0.10 & 0.07\pm 0.02 & 19.35\pm0.15\\
\noalign{\smallskip}

\mathrm{SMC\,3\_5} & 01\mathrm{:}27\mathrm{:}30.816 &
-74\mathrm{:}00\mathrm{:}49.320 & +8.9671 & 16 & 29/07/11 & 07/10/12 & 436 &
1.58\pm 0.06 & 1.02\pm 0.15 & 0.06\pm 0.01 & 19.34\pm0.22\\
\noalign{\smallskip}

\mathrm{SMC\,4\_2} & 00\mathrm{:}25\mathrm{:}14.088 &
-73\mathrm{:}01\mathrm{:}47.640 & -5.9198 & 15 & 18/10/12 & 23/06/14 & 606 &
1.57\pm 0.06 & 0.94\pm 0.08 & 0.05\pm 0.01 & 19.29\pm0.21\\
\noalign{\smallskip}

\mathrm{SMC\,4\_3} & 00\mathrm{:}45\mathrm{:}14.688 &
-73\mathrm{:}07\mathrm{:}11.280 & -1.1369 & 16 & 29/08/11 & 18/08/13 & 720 &
1.56\pm 0.05 & 0.92\pm 0.10 & 0.07\pm 0.01 & 19.27\pm0.17\\
\noalign{\smallskip}

\mathrm{SMC\,4\_4} & 01\mathrm{:}05\mathrm{:}19.272 &
-73\mathrm{:}05\mathrm{:}15.360 & +3.6627 & 15 & 25/09/12 & 14/07/14 & 648 &
1.56\pm 0.06 & 0.87\pm 0.06 & 0.07\pm 0.02 & 19.27\pm0.16\\
\noalign{\smallskip}

\mathrm{SMC\,4\_5} & 01\mathrm{:}25\mathrm{:}11.016 &
-72\mathrm{:}56\mathrm{:}02.040 & +8.4087 & 18 & 29/08/11 & 25/11/13 &819 &
1.56\pm 0.06 & 0.92\pm 0.17 & 0.06\pm 0.01 & 19.42\pm0.19\\
\noalign{\smallskip}

\mathrm{SMC\,5\_2} & 00\mathrm{:}26\mathrm{:}41.688 &
-71\mathrm{:}56\mathrm{:}35.880 & -5.5717 & 17 & 09/06/11 & 10/11/12 & 520 &
1.53\pm 0.06 & 0.99\pm 0.18 & 0.07\pm 0.01 & 19.24\pm0.27\\
\noalign{\smallskip}

\mathrm{SMC\,5\_3} & 00\mathrm{:}44\mathrm{:}49.032 &
-72\mathrm{:}01\mathrm{:}36.120 & -1.2392 & 19 & 04/10/12 & 08/08/14 & 673 &
1.54\pm 0.07 & 0.95\pm 0.12 & 0.06\pm 0.01 & 19.31\pm0.15\\
\noalign{\smallskip}

\mathrm{SMC\,5\_4} & 01\mathrm{:}04\mathrm{:}26.112 &
-71\mathrm{:}59\mathrm{:}51.000 & +3.4514 & 18 & 26/10/10 & 23/10/12 & 728 &
1.58\pm 0.08 & 0.92\pm 0.11 & 0.07\pm 0.01 & 19.35\pm0.15\\
\noalign{\smallskip}

\mathrm{SMC\,5\_6} & 01\mathrm{:}41\mathrm{:}28.800 &
-71\mathrm{:}35\mathrm{:}47.040 & 12.3004 & 18 & 30/09/11 & 11/09/14 & 1077 &
1.58\mathrm{\pm}0.08 & 1.0\mathrm{\pm} 0.2 & 0.05\mathrm{\pm} 0.01 & 19.15\pm0.65\\ 
\noalign{\smallskip}

\mathrm{SMC\,6\_3} & 00\mathrm{:}45\mathrm{:}48.768 &
-70\mathrm{:}56\mathrm{:}08.160 & -1.0016 & 14 & 20/08/11 & 06/11/13 & 809 &
1.52\pm 0.06 & 0.95\pm 0.12 & 0.05\pm 0.01 & 19.39\pm0.22\\
\noalign{\smallskip}

\mathrm{SMC\,6\_5} & 01\mathrm{:}21\mathrm{:}22.488 &
-70\mathrm{:}46\mathrm{:}10.920 & +7.5039 & 14 & 06/09/11 & 18/10/13 & 773 &
1.49\pm 0.03 & 0.97\pm 0.17 & 0.05\pm 0.01 & 19.42\pm0.17\\
\noalign{\smallskip}

\hline
\hline
\end{array}
\]
\end{table*}
However, as we are writing this paper, none of the available optical catalogues cover 
the field of view of VMC entirely (see Figure 4 of \citealt{Mor14}). There are external regions in both the LMC and SMC, and the whole
Bridge area where  a comprehensive census of all types of variable stars is still missing. 
On the other hand, in VMC each field is observed in the $K_\mathrm{s}$-band at least 13 times~\citep{Cio11}: 
11 times with an exposure of 750s (deep observations, hereafter) and twice with shorter exposures of 375s each  (shallow observations, hereafter)
 obtained over more than one year time-span. Furthermore,  the VSA provides a list of flags related to the analysis of the light 
curves of each VMC source \citep{Cro12}, that can be used to identify variable sources. 
In this paper we show that, in spite of the  
 small amplitudes in the near-infrared, pulsating variable stars can be effectively detected from the VMC data 
if a proper analysis is performed. Specifically, we combine
the information from:  i) the ($J-K_\mathrm{s}$, $K_\mathrm{s}$) colour-magnitude and the ($Y-J$, $Y-K_\mathrm{s}$) 
 colour-colour diagrams, ii) the analysis of the $K_\mathrm{s}$-band light curves,
iii) the Period-Luminosity relations, 
and iv) the VSA flags,  to devise a procedure and detect variable stars in the external regions of the Magellanic system
using only the VMC data.

This paper is organised as follows. The reliability of the VSA flags in detecting variable stars is presented in Section 2. 
Section 3 describes the method to identify variable stars in the SMC where information is available from the OGLE III survey.
In Section 4 we apply our method to identify classical Cepheids (CCs) 
 in external regions of the SMC
 where the VMC observations are complete and no other CC variable star catalogue is currently available. 
 Our results are summarized in Section 5.

\section{Identification of variable stars using the VSA flags}
We have tested the ability of the VSA in 
 identifying variable stars 
 based solely on the VMC data in three most external and completely observed tiles in the LMC, namely,  
tile LMC 6\_8, 7\_3 and 8\_8. The upper portion of Table~\ref{tab:qc} provides centre 
coordinates  of these 3 LMC tiles, together with information related to their observations.
We specifically selected these three tiles because they contain a large number of variable stars of different types 
and are located in external fields of the LMC with similar level of crowding as  in the regions 
where we currently have VMC data, but there is no coverage by the optical surveys.
We  first cross-matched the OGLE catalogues (\citealt{Sos08a,Sos08b,Sos09a,Sos09b,Sos12}; 
\citealt{Pol10}; \citealt{Gra11}) of variable stars in these fields against the VMC catalogue  
of the corresponding three tiles. We adopted a pairing radius of 0.5 arcsec in order to maximize the 
reliability of the matching procedure as discussed in \cite{Rip15}.
There are 7206 variable stars in common between the two catalogues; in particular we have 94 $\delta$ Scuti (DSCT) stars,
1071 RR Lyrae (RRLYR) stars, 1521 Eclipsing binaries  (ECL), 217 CCs, 4289 Long Period Variables (among wich are 83 Mira, 3626 OGLE Small
Amplitude Red Giants -- OSARGs, and 580 SemiRegular Variables -- SRVs), 6 Type II Cepheids (T2Ceps) and 8 Anomalous Cepheids (ACs).
For all of them we investigated the Variability classification provided by the VSA; this is summarized by the 
{\it ``Variability Class''} parameter which is the result of the analysis performed by the WFAU team on the 
$Y$, $J$ and $K_\mathrm{s}$-band light curves of the sources. {\it ``Variability Class''}  is either 1 or 0,  whether or not
 the object is classified as variable (\citealt{Cro09,Cro12}). Table~\ref{tab:var} compares the number of 
OGLE variable stars cross-matched with VMC sources  in the aforementioned tiles 
and the number of them classified as variables by the VSA ({\it VariableClass}=1).  
The effectiveness of the VSA in selecting variable stars depends 
 on the mean magnitude, period and amplitude of the light variation  and, in turn,
 on the type of variability.
Indeed, according to the numbers in Table~\ref{tab:var} the type of variable stars most easily classified as such by 
the VSA are the LPV-Mira, ACs, LPV-SRV, T2Cep's and CCs. Note that the sample of T2Ceps and of ACs are not statistically representative.
In this paper we focus on the detection of Cepheids, CCs  in particular, because they are numerous 
and will give us a very good statistical sample to work with. We leave for a future paper the analysis of other types of variables.
CCs typically have  single-epoch photometric errors in the $K_\mathrm{s}$-band of about 0.01 mag  \citep{Rip12b,Mor14},  
typical  $K_\mathrm{s}$ amplitudes between 0.05 and 0.38 mag and periods between $\sim$0.5 d and tens of days. 
The search for CCs was performed in  regions of the Magellanic system  where we expect this kind of variable stars to be present and 
the VMC observations are complete, but for which no optical data are currently available. 
In particular,  we focused on the external regions of the SMC (see Section~\ref{sec:using}).

\begin{table}
\caption{VSA classification of variable stars within VMC tiles LMC 6\_8, 7\_3 and 8\_8.
 Col. 1: Type of variable stars; Col. 2:
number of OGLE~III stars with a VMC counterpart within 0.5 arcsec; Col. 3: number of stars classified as
variables by the VSA; Col. 4: percentage value. See text for details.}
\label{tab:var}
\begin{tabular}{lrrr}
\hline
\hline
\noalign{\smallskip}
Type   & $N_{\rm OGLE}$ &  $N_{\rm VSA}$  & $N$\%  \\
\noalign{\smallskip}
\hline
\noalign{\smallskip}
\noalign{\smallskip}
DSCT         &      94    &     1    &   1\%     \\
\noalign{\smallskip}
RRLYR        &     1071   &    28    &   3\%     \\
\noalign{\smallskip}
ECL          &     1521   &   132    &   9\%     \\
\noalign{\smallskip}
LPV-OSARG    &     3626   &   576    &   16\%     \\
\noalign{\smallskip}
CC           &      217   &   137    &   63\%     \\
\noalign{\smallskip}
T2Cep        &        6   &     4    &   67\%	       \\
\noalign{\smallskip}
LPV-SRV      &      580   &   477    &   82\%     \\
\noalign{\smallskip}
AC           &        8   &	7    &	 87\%	  \\
\noalign{\smallskip}
LPV-Mira     &       83   &    75    &   90\%     \\
\hline
\end{tabular}
\\
\end{table}

\section{Defining the method to detect classical Cepheids}\label{sec:CCSMC}
\begin{figure}
\begin{centering}
\includegraphics[width=8cm]{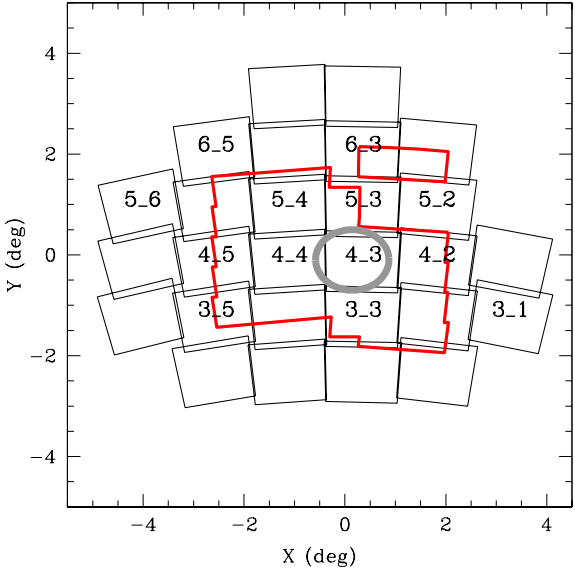} 
\caption{Celestial distribution of OGLE~III  footprint (red contours) and VMC tiles (black rectangles) in the SMC area. 
$X$ and $Y$ are defined as in van der Marel \& Cioni (2001) with $\alpha_0=12.5$ deg and $\delta_0=-73.0$ deg.
Tiles analysed in this work are labelled with their VMC IDs. The grey oval shows 
the area containing the peak density of the SMC CCs which was avoided in the present analysis.}
\label{fig:SMCtiles}
\end{centering}
\end{figure}

In order to specifically optimize the 
detection of  CCs, we first analyzed the properties of the VMC data for variable stars in the SMC 
classified as CCs by OGLE~III \citep{Sos10a}. These authors identified 4630 CCs in the SMC, of which 2626 
are fundamental-mode (F), 1644 are first-overtone (1O), 83 are second-overtone (2O), 274 are double-mode 
(59 F/1O and  215 1O/2O) and three are triple-mode CCs. 
 The mean $I$-band magnitude of these CCs ranges from $\sim$10 mag to $\sim$19 mag, with two major 
 peaks  at $I\sim$16.8 mag and $I\sim$15.8 mag.
We have used the OGLE~III coordinates of the SMC CCs as reference in the following analysis.
Figure~\ref{fig:SMCtiles} shows the distribution of the OGLE~III footprint (red contours) 
and the VMC tiles (black rectangles)  in the SMC area. 
The SMC tiles analysed in this work are the ones for which the VMC observations were completed before 2014 September 30. 
They are tiles SMC 3\_3, 3\_5, 4\_2, 4\_3, 4\_4, 
4\_5, 5\_2, 5\_3, 5\_4, 6\_3 and 6\_5, labelled with their IDs in Fig.~\ref{fig:SMCtiles}.
The lower portion of Table~\ref{tab:qc} lists centre coordinates and observation properties 
for all of them. 
OGLE~III observations cover the most central and crowded regions of the SMC. 
In particular, the peak in stellar density  (see Figure 1 of~\citealt{Rub15}) 
and  CC density (see right panel of  Figure 18 in~\citealt{Mor14})  in the SMC occurs 
in  an area centered at coordinates RA=12 deg, Dec=$-$73.1 deg and radius of 2
deg in RA and 0.5 deg in Dec. This region, marked by a grey oval in Fig.~\ref{fig:SMCtiles}, 
mainly covers the tile SMC 4\_3 and contains 1695
 of the 4630 CCs identified by the OGLE~III survey in  the SMC.
However, as we aim at fine-tuning our procedures to identify new SMC CCs outside the OGLE~III field (see Section 4), we considered in the following
 analysis only CCs lying in the external portions of the OGLE~III footprint, where the level of  crowding and reddening is similar to what we found 
 in the regions which currently have only VMC data (see Fig.~\ref{fig:SMCtiles}). Specifically, we considered 2935 CCs located  outside the oval shape in Fig.~\ref{fig:SMCtiles}.
By matching the OGLE~III and VMC catalogues, using a  pairing radius of 0.5 arcsec 
we obtained a sample of 2411 CCs that have VMC $K_\mathrm{s}$-band light curves. 
 Increasing the paring radius to 1.0 arcsec would result in
increasing by about 1\% the number of misidentifications. Conversely, reducing the paring radius to
0.1 arcsec, would result in loosing about 10\% of the crossmatched sources. On the other hand, the
reliability of our cross-identifications and, in turn, of the value adopted for the paring radius, is
specifically assessed by folding the VMC $K_\mathrm{s}$-band light curves according to the corresponding
OGLE III periods (e.g. \citealt{Rip15}).
These CCs lie in the 
 tiles (see Fig.~\ref{fig:SMCtiles}): SMC 3\_3, 3\_5, 4\_2, 4\_3 and 4\_4 (excluding sources inside the oval grey contour), 4\_5, 5\_2 (two sources),
 5\_3, 5\_4, 6\_3 (four sources) and 6\_5 (two sources). 

\subsection{Selection based on the colour-magnitude and colour-colour diagrams}\label{sec:cmd-cc-sel}
\begin{figure*}
\begin{centering}
\includegraphics[width=8cm]{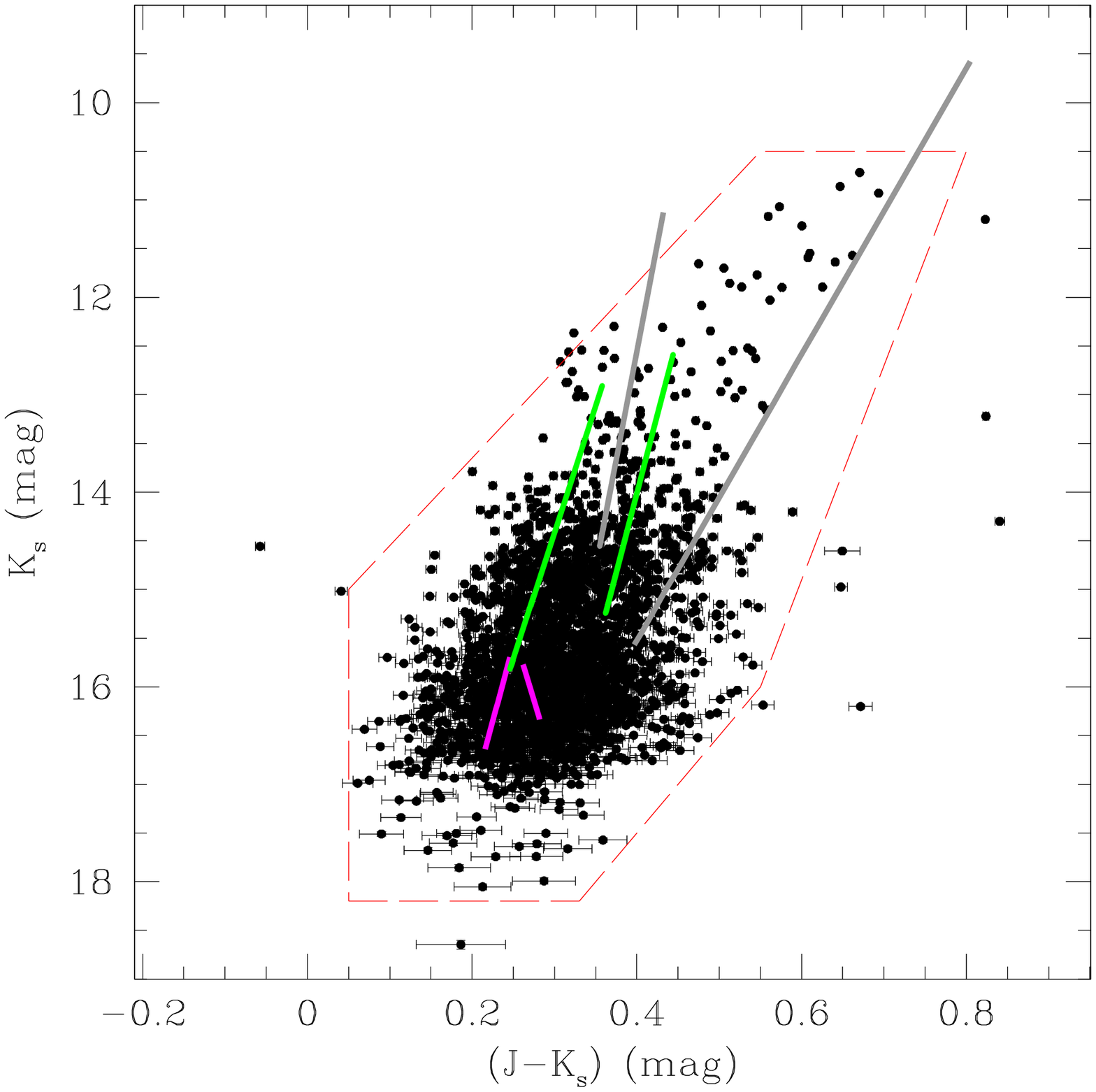} 
\includegraphics[width=8cm]{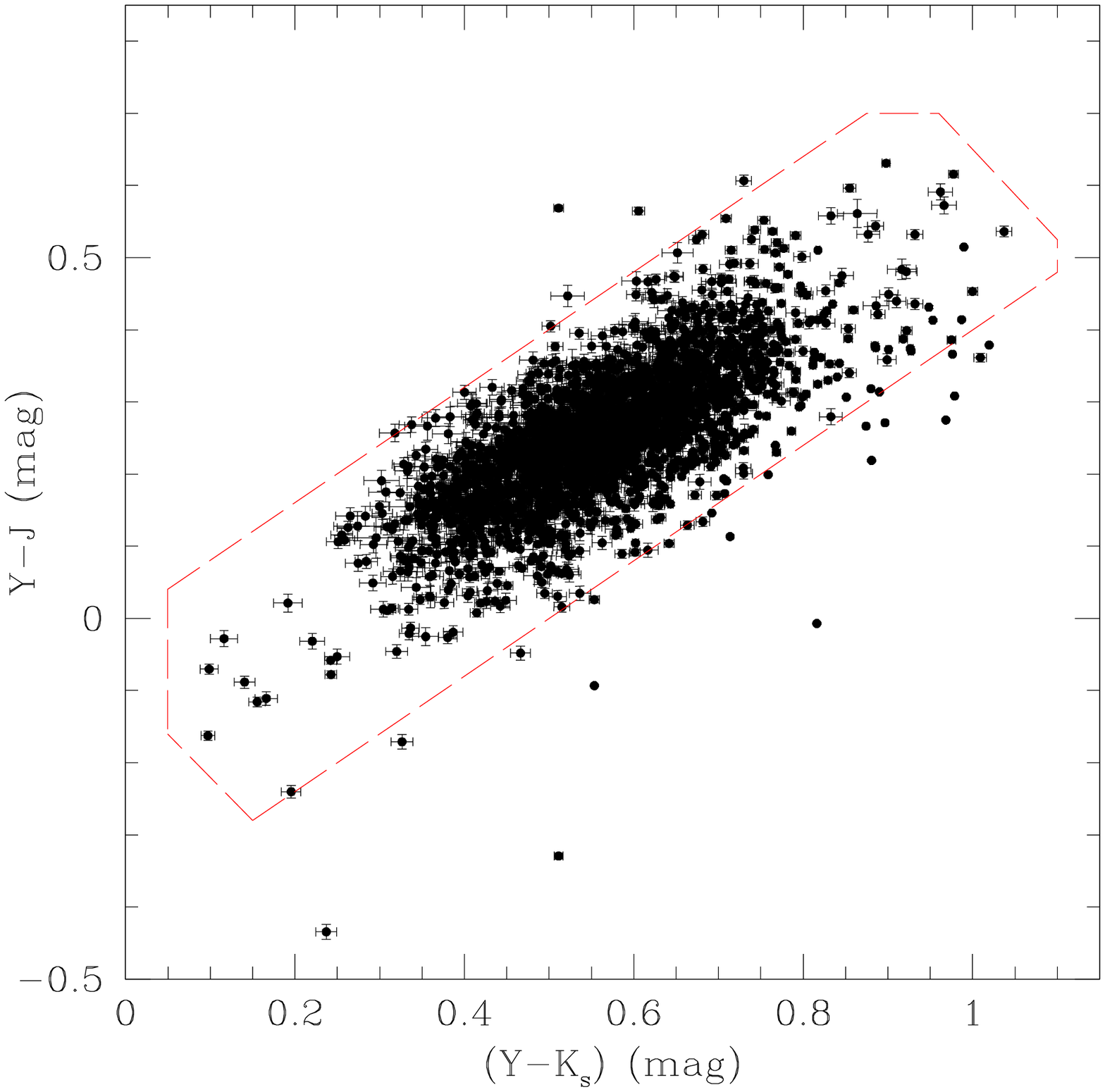}
\caption{ Left: Distribution of known SMC CCs in the ($J-K_\mathrm{s}$, $K_\mathrm{s}$) colour-magnitude diagram. 
Grey, green and magenta solid lines represent the theoretical 
 blue and red edges of the CC ISs for F, 1O and 2O pulsation modes, respectively. 
Right: Same as the left panel but  in the ($Y-J$, $Y-K_\mathrm{s}$) colour-colour diagram.
In both panels, red dashed contours indicate the boundaries of the region that will be used to select candidate CCs in the external part of the SMC.
See Section~\ref{sec:CCSMC} for details.}
\label{fig:cmdKJK}
\end{centering}
\end{figure*}

We used the sample of 2411 OGLE~III CCs described in the previous section  as a reference to define the range in VMC colours and magnitudes in which the SMC CCs lie. 
Figure~\ref{fig:cmdKJK} shows the distribution of such stars in the ($J-K_\mathrm{s}$, $K_\mathrm{s}$) 
colour-magnitude diagram. 
We have highlighted with red dashed contours the region that we
will use to select CC candidates. We note that 2365 of the 2411 reference CCs (corresponding to 98\% of the sample) have 
$\langle K_\mathrm{s} \rangle <$ 17.0 mag and only 2\% of the population have fainter magnitudes. Specifically, 45 stars 
have       $17.0 \leq \langle K_\mathrm{s} \rangle \leq$ 18.2 mag and only 1 CC is as faint as $\langle K_\mathrm{s} \rangle$ = 18.6 mag. 
We have also reported in Fig.~\ref{fig:cmdKJK} the theoretical instability strips (ISs) for CCs with metallicity 
{\it Z}=0.004 and helium abundance $Y$=0.25 taken from \cite{Bono00,Bono01a,Bono01b}. These values of $Z$ and $Y$ are 
appropriate for the SMC CCs.  To transform  the theoretical IS edges to the observational plane we adopted 
the static model atmospheres by \citet{CGK97a,CGK97b}, 
an absorption $A_V$=0.1 mag \citep{Has11} 
and a distance modulus of 19.0 mag, computed as the weighted average of the CC results in 
table 1 of \cite{Has12}\footnote{This method applied to 
 the CC results listed in Table 2 of \cite{deG15} leads exactly to the same distance modulus value.}. 

Specifically,  we have plotted in Fig.~\ref{fig:cmdKJK}  the blue and red edges of the IS for CCs of different pulsation modes: F (grey
lines), 1O (green lines) and 2O (magenta lines). 
 For magnitudes brighter than 14 mag, there is reasonably good agreement, within the errors, between theoretical and observed IS's, 
suggesting that the sample of CCs we are using as reference \citep{Sos10a}, once matched with the 
VMC data, represents the ranges in $K_\mathrm{s}$ magnitude and $J$-$K_\mathrm{s}$ colour  covered by the SMC CCs very well. 
A number of reference CCs, especially at magnitudes fainter than $K_\mathrm{s} \sim$14 mag, are found beyond the boundaries 
of the theoretical IS's, both at bluer and redder colours.
 This is likely caused by 
metallicity, differential reddening issues and by the poor sampling of their VMC $J$-band light curves. 
Indeed, the same theoretical  ISs are in much better agreement  with other 
observational samples, including  Magellanic Cepheids, where light curves are better sampled 
in colour (e.g. \citealt{Bon99}, \citealt{Mar05}).
 We also compared the colour-magnitude distribution for sources with more than 10 data points in the
$J$ band (about 10\% of our sample) using the $J$-$K_\mathrm{s}$ colours 
obtained after analyzing both the $J$ and $K_\mathrm{s}$-band light curves 
with a custom template fitting procedure \citep{Rip15,Rip16}. We obtained a relatively 
more compact distribution, confirming that the 
poor sampling of the VMC $J$-band light curves is one of the possible issues. 

The right panel of Figure~\ref{fig:cmdKJK} shows the distribution of these CCs in the 
$Y-J$, $Y-K_\mathrm{s}$  colour-colour diagram. 
As in Fig.~\ref{fig:cmdKJK},  we have highlighted with red dashed contours the region that 
we will use in Sec.~\ref{sec:using} to select CC candidates. 

\subsection{Analysis of the light curves}\label{sec:lc-analysis}
 OGLE~III periods of the reference CCs range between 0.25 d to 208 d. 
  The vast majority of the reference CCs (2240 sources) have $P$$<$5d, 109 sources have
 $5\leq$$P$$<$10 d, 60 have $10\leq$$P$$<$75 d and only 2 sources have $P$$\geq$75 d. Specifically, these two sources have 
 $P_{\rm OGLE}$= 128.2 d and $P_{\rm OGLE}$ =208.8 d; despite their $K_\mathrm{s}$-band photometry ($K_\mathrm{s} \sim$ 11 mag) 
 should not be saturated, they are fainter than expected according to the $K_\mathrm{s}$-band $PL$ relation, especially the last one. 
 The same feature occurs in the $I$-band OGLE data. 
Despite their images do not show any clear problem, their folded $K_\mathrm{s}$-band light curves 
 are very noisy without a clear shape; for these reasons we decided to discard them and focus on the sources
with $P$ lower than 80 d. 

We analysed the $K_\mathrm{s}$-band  light curves 
 and derived the period of the reference CCs 
from the $K_\mathrm{s}$ time-series data alone,  ignoring OGLE~III  information on the period. 
We used all available VMC epochs to study the light curves including observations obtained during 
nights with sky conditions (i.e. seeing and ellipticity) that exceeded the VMC requirements \citep{Cio11},
since our fitting procedure is able to handle lower accuracy data (see, e.g., \citealt{Rip15} and references therein).
The  resulting light curves have a number of data points that ranges from  7 to 60.  
We checked the images of some CCs with a  few epochs and found that often these sources are contaminated by very bright companions.

  Periods were derived  using the program ``Significance Spectrum'' ({\it SigSpec}; \citealt{Ree11}).
  {\it SigSpec} is a method specifically developed for detecting and characterizing periodic signals in
noisy data. While most period search analyses explore only the Fourier amplitude, through the power spectrum, ignoring phase information,  {\it SigSpec}  
is based on the definition of a quantity called {\it spectral significance} for a time series, a function of Fourier phase and amplitude. 
The {\it spectral significance} quantity conveys more information than does the conventional amplitude spectrum alone, and appears to simplify 
statistical issues as well as the interpretation of phase information.
  
\begin{figure}
\begin{centering}
\includegraphics[width=8cm]{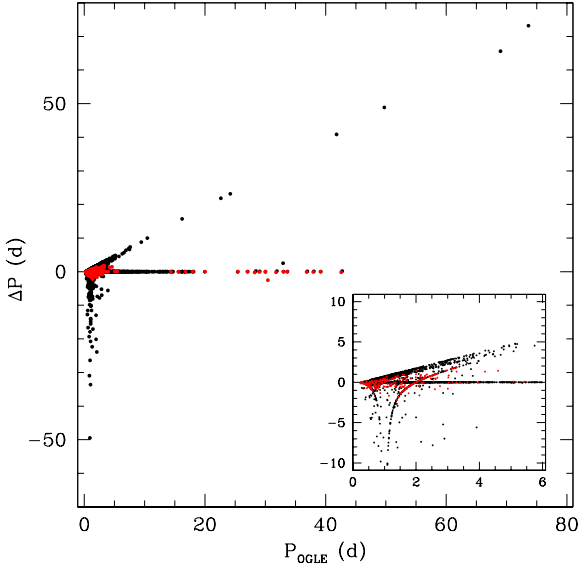}
\caption{Distribution of $\Delta P$=$P_{\rm OGLE}-P_{\rm SigSpec}$ versus $P_{\rm OGLE}$. Shown in the inset
is an enlargement of the short period region. Red points indicate stars within 3 $\sigma$ from  
the $PL$ relations but with $\Delta P$ larger  than 0.02 d. See text and Fig.~\ref{fig:plSMCknown} for details.}
\label{fig:PdeltaP}
\end{centering}
\end{figure}

We ran {\it SigSpec} on the VMC  $K_\mathrm{s}$-band 
  time-series  adopting a lower period of 0.25 d,   an upper period of 80 d
  and weighting by the $K_\mathrm{s}$ single-epoch errors.   
  The r.m.s. of the light curve analysis typically ranged 
 from 0.002 mag  to 0.15 mag, (with a few extreme values as  large as  0.3 mag) and a median value of 0.015 mag. 
The median value of the {\it spectral significance} is $\sim$ 3.0 
with a standard deviation of 0.9, and minimum and maximum values of 0.9 and 7.0 (see also discussion at the end of Section~\ref{sec:PL-selection}). 
Figure~\ref{fig:PdeltaP} shows the comparison between the OGLE~III periods and
the periods we derived  for the reference CCs running {\it SigSpec} on the $K_\mathrm{s}$ time-series data.
 For 54\% of the sources (1302 stars)  the two periods are in good agreement, the difference being smaller 
than 0.02 d.  On the other hand, 
for some stars the period found by {\it SigSpec} ($P_{\rm SigSpec}$) is definitely shorter than that published by  the OGLE team.  
In particular,  about  800 sources have $\Delta P$=$P_{\rm OGLE}$ (where $\Delta P$=$P_{\rm OGLE}-P_{\rm SigSpec}$) 
because their $P_{\rm SigSpec}$  is near zero (see Fig~\ref{fig:PdeltaP}). 
This is probably due to alias problems in the case of sources with period shorter than a few days and to saturation 
of the $K_\mathrm{s}$ time series for stars with longer period. Conversely, there are about 150 stars for which $P_{\rm SigSpec}$  
is definitely longer  than $P_{\rm OGLE}$ and hence $\Delta P$ assumes large negative values (see Fig~\ref{fig:PdeltaP}). This might be
due to faintness and hence poor quality of the $K_\mathrm{s}$ light curves of these stars that affect the period search procedure.
We also checked if there is any   dependence of $\Delta P$ on the amplitude of the light curves, but did not find any.
We do not have an estimate of the error on $P_{\rm SigSpec}$ for each star but, assuming the OGLE as the correct one, 
we can use as an estimate of the $P_{\rm SigSpec}$ error the median value of $\Delta P$, that is 0.002 d.

As an additional test, we performed an analysis of the $K_\mathrm{s}$ light curves using a different period search program,
 FNPEAKS \citep{Kur85}. We adopted the same limits (0.25-80 d) for the period and a frequency step of 0.0001 s$^{-1}$.
For 29\% of the stars there is good agreement between the OGLE~IIII and the FNPEAKS periods:  
$\arrowvert \Delta$P$\arrowvert \leq$ 0.02 d. FNPEAKS  does not allow to 
weight by the error of the single epoch magnitudes; this probably explains the lower percentage of periods recovered within 
0.02 d with respect to {\it SigSpec} (54\%). We also checked the $P$ versus $\Delta P$ plot,  that is the counterpart of Fig.~\ref{fig:PdeltaP}. 
The shape of the $\Delta P$=$P_{\rm OGLE}-$$P_{\rm FNPEAKS}$ versus  $P_{\rm OGLE}$ is very similar to Fig.~\ref{fig:PdeltaP}, hence confirming that  FNPEAKS and {\it SigSpec}  find consistent 
results, however,  {\it SigSpec} appears to be more efficient. 
In the analysis described in Sec.~\ref{sec:using} we will thus adopt {\it SigSpec} to estimate the period of the candidate variable stars.

\subsection{Selection based on the Period-luminosity relation}\label{sec:PL-selection} 
\begin{figure}
\begin{centering}
\includegraphics[width=8cm]{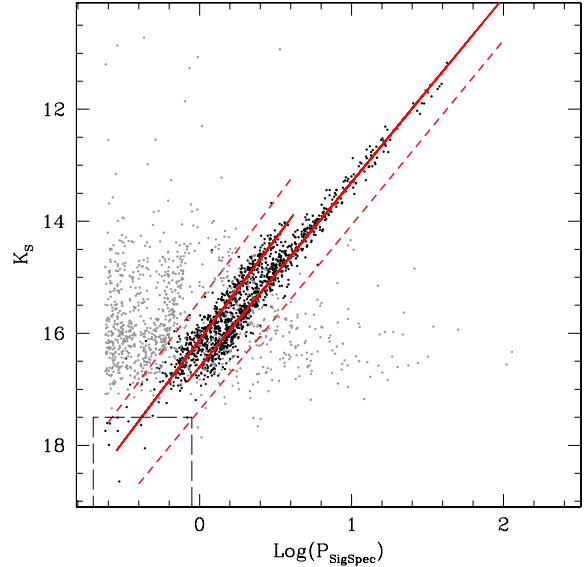}
\caption{Distribution of known SMC CCs in the Period-$K_\mathrm{s}$-band Luminosity plane 
using for the period values derived from the analysis of the
$K_\mathrm{s}$-band light curves.  
Red solid lines show the $PL$ relations for F (right) and 1O  (left) CCs derived using the OGLE~III periods. 
Red dashed  lines represent 3$\sigma$ boundaries of the 1O (upper line) and F (lower line)  $PL$s, respectively. Black and grey dots show stars 
with $\arrowvert \Delta$P$\arrowvert \leq$0.02 d  and  $\arrowvert \Delta$P$\arrowvert >$ 0.02 d, respectively.
A black dashed box marks the region populated by RR Lyrae stars. See Section~\ref{sec:contamination} for details.}
\label{fig:plSMCknown}
\end{centering}
\end{figure}

Figure~\ref{fig:plSMCknown} shows the  $K_\mathrm{s}$-band period-luminosity 
($PL$) plane obtained for the 2411 reference CCs using for the period values derived with the {\it SigSpec} analysis 
 and for the  $K_\mathrm{s}$ average magnitude values available from the VSA. 
Red solid lines show the $PL$ fits obtained for F and 1O mode CCs, 
using the periods from OGLE~III;  red dashed lines show the 
corresponding 3$\sigma$ boundaries for 1O (upper dashed line) and  F (lower dashed line) CCs, respectively. 
 As expected, Cepheids with a good {\it SigSpec} estimate of the period ($\arrowvert \Delta$P$\arrowvert \leq$0.02 d, black points) 
 lie near the $PL$s defined using the OGLE~III periods.
Specifically, there are  1589 objects lying within 3$\sigma$  from the $PL$s, corresponding to 66\% of the original sample\footnote{This number reduces to 1139, corresponding to
47\% of the original sample, if the periods obtained with FNPEAKS are used instead.}. This number includes  1304 stars
with $\arrowvert \Delta$P$\arrowvert \leq$ 0.02 d (black points) and 285 sources with $\arrowvert \Delta$P$\arrowvert$ larger than 0.02 d (grey points). 
These 285 sources are marked by red points in the  $ \Delta$P versus $P_{\rm OGLE}$ distribution shown in Fig.~\ref{fig:PdeltaP}. 
The bulk of the distribution  has  $\arrowvert \Delta$P$\arrowvert \leq$ 1 d  (253 objects), and only a few 
sources (32 stars) have $ \Delta$P values larger than $\pm$ 1 d and up to 2.5 d.

Stars within the 3$\sigma$ from the $PL$s, show an average {\it significance} (see Section~\ref{sec:lc-analysis}) of 3.4 with minimum and 
maximum values of 1.8 and 7.0,  respectively. 
We  checked the position of these stars on the $PL$ according to their $significance$
value and  noted that several stars with {\it significance} value between 1.8 and 2.0 lie within 2$\sigma$ from the $PL$s.
Hence, in the following analysis we will retain only sources with {\it significance} larger than 1.7 (Section~\ref{sec:using}).
  We will also take into account the possible contamination by other types of variable stars. 
In particular, the RR Lyrae stars  follow a $PL_{K_\mathrm{s}}$ relation that, although different, partially overlaps 
with the $PL_{K_\mathrm{s}}$ relation of CCs.  A black dashed box schematically shows the locus of the SMC RR Lyrae stars
in the  $K_\mathrm{s}$-band $PL$ plane in Fig.~\ref{fig:plSMCknown}. This issue will be discussed in more detail 
in Section~\ref{sec:contamination}. 

\subsection{ VSA flags}\label{sec:VSA-selection} 
The VSA provides several flags describing  the quality of the  light curve of each VMC source.
A  complete explanation of these parameters is provided in \cite{Cro12} and on the VSA web page\footnote{http://horus.roe.ac.uk/vsa}.  
Here we briefly  summarize the properties that are relevant for  the present study. 
The VSA classifies sources according to their nature by the $mergedClass$ parameter within the $vmcSource$ table,
 containing the information about the sources extracted from the stacked images. 
Specifically,  the association between parameter value and physical nature of
the source is as follows: 1=galaxy, 0=noise, $-1$=stellar, $-2$=probableStar, $-3$=probableGalaxy, $-9$=saturated source.
 The $KsppErrBits$ parameter encodes quality issues associated with a given $K_\mathrm{s}$-band detection within the vmcSource table. 
Its value is zero for a detection without quality issues and grows according to the severity of the issue. 
In particular to include sources with only minor $K_\mathrm{s}$-band quality issues the user can filter as $KsppErrBits<$256. 

We used the VSA flags to select among our sample of 2411 reference CCs only stars with at least 10 data points (this corresponds to 2407 of the 2411 sources),  
that are classified as stars or probable stars by the VSA ($mergedClass$=$-$1 or $-$2), as variables 
($VarClass$=1) and that do not exibit any severe quality issues
 ($KsppErrBits<$256),  thus ending up with a total of  1445 sources. This corresponds to 60\% of our original sample. Five hundred  
  of the sources that the VSA classifies as non variable sources ($VarClass$=0) perfectly lie within the 3$\sigma$ limits from the $PL$s. Hence, 
 the VSA flags, although useful for a first selection of variable sources, need to be fine-tuned (see, e.g., \citealt{FLC15}) to increase
 their capability to detect bona fide variable stars.
 In the next section we will use the VSA flags to corroborate our identification of new SMC CCs based on the
 colour-magnitude and colour-colour diagrams and $PL$ selections
 described in the previous sections.

In conclusion, to  identify CCs from the VMC $K_\mathrm{s}$-band time series data 
we will 
\begin{itemize}
\item first select candidate variables by applying the colour-magnitude and colour-colour diagrams cuts described in Section~\ref{sec:cmd-cc-sel}, 
\item analyse with {\it SigSpec} the light curves 
of the candidate variables to  determine their periods (see Section~\ref{sec:lc-analysis}),
\item consider as best candidates  the stars falling within 3$\sigma$ from the $PL$s defined as described in Sec.~\ref{sec:PL-selection}, 
\item investigate the VSA flags. 
\end{itemize}
In this way, following a visual inspection of the light-curves to identify bona fide CCs, we should be able to identify 66\% of the CCs that populate the 
external areas of the SMC analysed in the next Section.

\section{Detection of classical Cepheids in the external regions of the SMC}\label{sec:using}
\begin{figure}
\begin{centering}
\includegraphics[width=8cm]{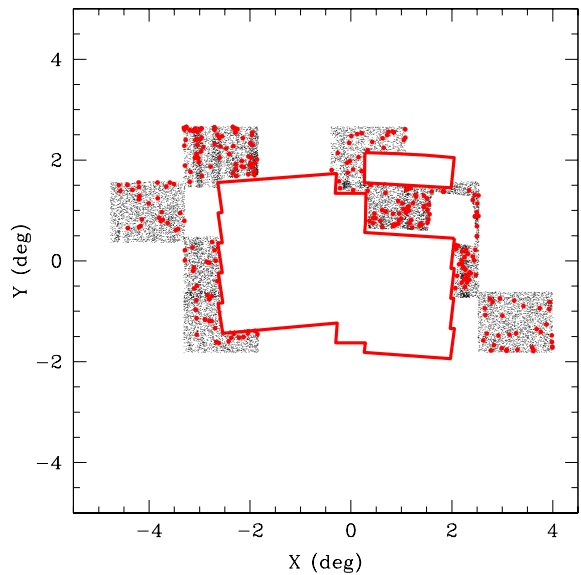} 
\caption{Spatial distribution of the SMC CC candidates (black points)  selected using the tools described in 
Section~\ref{sec:CCSMC}.
$X$ and $Y$, the red contours and the  VMC tiles are defined as in Fig.~\ref{fig:SMCtiles}.
The white area within tile SMC 5\_2 at $X\sim 2$, $Y\sim1$ corresponds to  the 47 Tucanae field analyzed by Weldrake et al. (2004). 
Sources that we confirm to be bona fide new Cepheids are marked by red filled circles (see  text for details).}
\label{fig:SMCfield}
\end{centering}
\end{figure}
We have applied the methods described in Section~\ref{sec:CCSMC} to look  for CCs in all regions of the SMC where
we currently have complete VMC data, but where no optical comprehensive catalogue of variable stars is available yet. 
Figure~\ref{fig:SMCfield} shows the portion of sky
we have analysed with the different VMC tiles labelled in Fig.~\ref{fig:SMCtiles} (see also Tab.~\ref{tab:qc}).  
We specifically considered only regions outside the OGLE~III footprint  
(red solid lines) and also discarded the region studied by Weldrake et al. (2004; empty area within tile SMC 5\_2
at $X\sim 2$, $Y\sim1$),  
who provided a comprehensive catalogue of variable stars in the field of the Galactic globular cluster  47 Tucanae. 
In particular, we first selected our candidates in the aforementioned VMC tiles,
and then discarded sources 
lying in the region that overlaps with the OGLE~III and \cite{Wel04} fields. The sources 
were then further selected using the colour-magnitude and colour-colour diagrams, as  
described in  Section~\ref{sec:cmd-cc-sel}, yielding 
19,938 candidate CCs, that  are plotted as black points in Fig.~\ref{fig:SMCfield}.

\begin{figure}
\begin{centering}
\includegraphics[width=8cm]{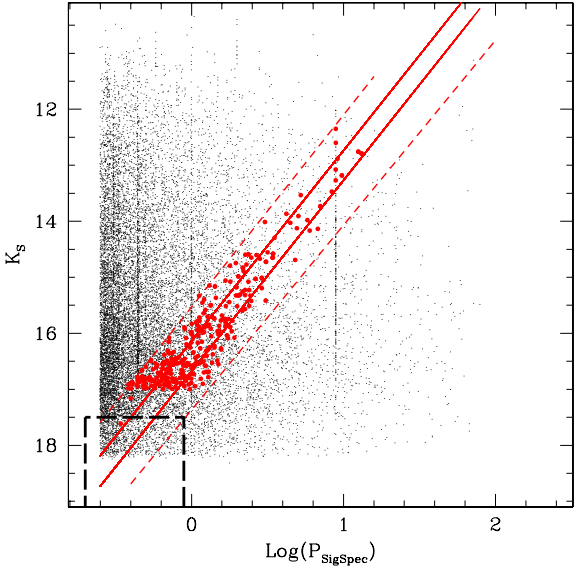} 
\caption{Distribution of the 18,090 candidate CCs (black points) in the Period - $K_\mathrm{s}$-band Luminosity plane.
Red solid lines show the  $PL$ relations and their 3$\sigma$ boundaries (red dashed lines) defined by the OGLE~III SMC CCs. 
 A black  dashed box highlights the region occupied
by the SMC  RR Lyrae stars. Red filled circles mark bona fide CCs identified in the present study. 
Their $K_\mathrm{s}$-band light curves are shown  in Fig.~\ref{fig:lightCurves1}.}
\label{fig:pl}
\end{centering}
\end{figure}

 \cite{Rub15} estimated the star formation history (SFH) in these regions of the SMC. By comparing the best-fitting SFH models 
 with the theoretical CC instability strips for  $Z$=0.004 and $Y$=0.25  
 we predict a few hundreds CCs to populate the area under 
investigation. Hence, clearly, only a few percent of our 19,938 CC candidates are bona fide CCs. 
For 18,090 of these sources the VSA provides $K_\mathrm{s}$-band light curves with at least 10 data points, we  
analysed them with {\it SigSpec} and defined their period following the procedure described in Section~\ref{sec:lc-analysis}. 
 Furthermore,  we derived average $K_\mathrm{s}$ magnitudes and $K_\mathrm{s}$-band  amplitudes (Amp$K_\mathrm{s}$) 
 analysing the light curves with an automatic template fitting procedure specifically developed for the  
analysis of the $K_\mathrm{s}$-band light curves of the SMC CCs \citep{Rip16}. 
We then used the periods derived with {\it SigSpec} and the average  $K_\mathrm{s}$
magnitudes computed as described above to plot the sources in the Period - $K_\mathrm{s}$ Luminosity 
plane shown in Fig.~\ref{fig:pl}.  
We have also plotted in  this figure 
 the 3$\sigma$ boundaries of the $K_\mathrm{s}$-band $PL$ relations defined by 
the OGLE~III  SMC CCs  (see Section~\ref{sec:PL-selection}).
 Excluding objects
more than 3$\sigma$ away from the F and 1O mode $PL$ relations the sample reduces to 4,817 sources.

We further selected the sample using the parameters of the template fitting procedure. 
A comprehensive description of this procedure and of its parameters is provided by \cite{Rip16}.
 Here we simply note that we used the parameter 
$G$, which gives an estimate of the goodness of the fit by weighting the residuals 
of the fit (rms) with the number of data points rejected (outliers) from the fitting 
procedure. The first term of $G$ tends to favour templates which give the smallest rms values, 
whereas the second term favours those removing the least number of outliers. 
The balance between these two terms generally yields an automatic selection of the best 
templates that is in agreement with the visual inspection of the fitting procedure.
Since $G$ is inversely proportional to the square of the rms and directly 
proportional to the fourth power of the ratio between used in the fit and total number 
of phase points in the light curve, values of $G$ in the interval 100-10000 generally 
mean good fit, and, in absolute terms, also good light curve with small scatter. On the contrary, 
values of $G$ below 100 are usually associated with highly scattered  light curves. 
Finally, we retained only sources with the $G$ parameter between 100 and 10,000 and  $K_\mathrm{s}$-band 
 amplitude $\geq$ 0.04  mag further 
reducing the sample to 3,636 best candidates with $\langle K_\mathrm{s} \rangle$ magnitude in the range of  12.31 to 18.21 mag.  This number appears to be still rather large if compared
to expectations from the SFH recovery.
A check was made by comparing the distributions of CCs in the VMC tiles partially covered by OGLE~III. 
In particular, half of the tile SMC 4\_5 is covered by both OGLE III and VMC and the other half only by VMC 
(see Figs.~\ref{fig:SMCtiles} and \ref{fig:SMCfield}). 
We divided the tile into four subregions, each of them approximately covering the same area. 
The two western subregions lie within the 
OGLE~III footprint, while the  two eastern ones lie outside. 
OGLE~III detected 53 CCs in the north-western subregion and 30 in the southern one. 
From the SFH performed by \cite{Rub15} and the theoretical IS's (\citealt{Bono00,Bono01a,Bono01b}), we obtain a total number of
about  55  (with a minimum of 35 and a maximum of 76) CCs from the SFH recovery, in good agreement with the OGLE~III findings. 
Our number of CC candidates in the two eastern subregions of tile SMC 4\_5 is roughly 10 times that observed by OGLE~III in the two western ones.
This test  suggests that the majority of the 3636 new candidate variable stars are not CCs and that  a high level 
of contamination must be present among them. In particular there could be contamination by other types of variables, but also issues such as: 
saturation (on the bright side),  limiting magnitude and photometric error problems (on the faint side),  blending effects, intrinsic problems of the NIR data, etc. 
These issues  may particularly affect faint sources, thus leading to an overestimate of the number of variable sources. 
 Indeed, the sample of 3,636 new candidate variables includes 
1,677 sources with $\langle K_\mathrm{s} \rangle <$ 17.0 mag and 1,959 fainter sources with  $17.0 \leq \langle K_\mathrm{s} \rangle \leq $ 18.2 mag. However, 
according to the discussion in Section~\ref{sec:cmd-cc-sel}, only 2\% of  the SMC CCs have  $\langle K_\mathrm{s} \rangle$ magnitudes fainter  than 17.0 mag. Hence,
the 1,959 candidate variables fainter than 17.0 mag should include, at most,  $\sim$ 40 bona fide CCs.
 We hence visually inspected only the $K_\mathrm{s}$ light curves of all 1,677 sources brighter than $\langle K_\mathrm{s} \rangle $= 17.0 mag 
 and selected among them 297  sources whose $K_\mathrm{s}$-band light curves have the typical shapes  of CCs.
 In particular, all sources with very poor light curve coverage were discarded since a firm 
classification was not possible. Moreover, for several sources the light curves did not show the typical shape of
cepheids, these sources were discarded as well. During the visual inspection, a percentage reliability flag has been 
assigned to each source. This flag is 100\% for a source with (i) good phase coverage, 
(ii) good light curve shape and (iii) good position in the PL, 
while its value diminishes according to issues related to one or more of the aforementioned features, becoming 62\% for sources that are not
confirmed to vary. 

 Then we also checked  a subsample of 240 among  the 1,959 sources fainter than 
 $\langle K_\mathrm{s} \rangle $ = 17.0 mag; this  provided two additional sources with light curves typical of CCs.
Particular attention was devoted to sources showing a $K_\mathrm{s}$-band amplitude between 0.04 mag and 0.1 mag since
at this level of amplitude it is hard to distinguish between real variable sources and spurious objects. We hence decided
to keep only low amplitude sources with a good light curve coverage and very clear shapes.

We consider this total sample of 299 candidate variable stars that passed  the 
colour-magnitude and colour-colour diagrams, $PL$, template 
parameters selections and the visual inspection of the light curve as bona fide Cepheids. 
 Out of 299, nine sources are in common with the General Catalog of Variable Stars (GCVS, \citealt{Art96}), 
two (VMC-SMC-CEP-258 and 286) are ACs in common with \citealt{Sos15a}, 
while 288 are new CCs identified in the present study.
 The new CCs have periods in the range 
from about 0.34 to 9.1 d and span the magnitude range: 12.9 $\leq \langle K_\mathrm{s} \rangle \leq$ 17.6 mag, 
and only two being fainter than 17.0 mag (see above). 
This number is well  consistent with the predictions of \cite{Rub15} SFH recovery  in these external regions of the SMC.  
In particular,  of the  299 confirmed Cepheids, only 13 
are located in the eastern subregions of  tile SMC 4\_5. In the same two subregions \cite{Rub15} SFH recovery leads to  a total number of
about  10  (with a minimum of 4 and a maximum of 20) CCs, in very good agreement with our findings.

Table~\ref{tab:CCcandidates1} lists our 299 
 SMC Cepheids along with their RA, Dec (J2000) coordinates, 
the period  obtained with {\it SigSpec} (for a discussion about the errors see Section~\ref{sec:lc-analysis}), 
the $K_\mathrm{s}$-band  amplitude and the intensity-averaged 
$K_\mathrm{s}$ magnitude computed with the template-fitting procedure, the {\it VarClass} 
parameter attributed by the VSA and the variability type ({\it VarType}) assigned in the present study 
together with comments, if any, and the aforementioned percentage reliability flag. The last column of  Table~\ref{tab:CCcandidates1} also provides the 
GCVS identification number, whenever appropriate. 
The sources are ordered by increasing magnitude going from brighter to fainter objects.  
Their spatial distribution over the SMC is shown in Fig.~\ref{fig:SMCfield} where  they have been marked as red filled circles.
\begin{table*}
\caption{Information on the 299 Cepheids that  we have  identified in the SMC using only the near-infrared photometry of the VMC survey.  
Col. 1:  VMC-SMC-CEP ID, Col. 2: Right Ascension (RA, J2000), Col. 3: Declination (Dec, J2000), 
Col. 4:  number of  $K_\mathrm{s}$ observations, Col. 5: period computed with {\it SigSpec}, 
Col. 6: $K_\mathrm{s}$ amplitude computed with the template fitting procedure, 
Col. 7:  intensity-averaged $K_\mathrm{s}$ magnitude  computed with the template fitting procedure,
Col. 8: {\it VarClass} flag assigned by the VSA,  Col. 9: variability type 
assigned in the present study, corresponding percentage reliability flag and specific comments, if any. The table is published in its entirety as 
Supporting Information with the electronic version of the article. 
A portion is shown here for guidance regarding its form and content.}
\label{tab:CCcandidates1}
\begin{tabular}{ccccccccl}
\hline
\hline
 VMC ID       &     RA        &        Dec       & $N_{K_\mathrm{s}}$& $P_{\rm SigSpec}$ &  Amp$K_\mathrm{s}$  & $\langle K_\mathrm{s} \rangle$ & {\it VarClass} &   {\it VarType \& Comments}\\  
             &    (deg)      &       (deg)      &                   &           (d)     &    (mag)            &    (mag)                       &                &                   \\
\hline
 001   &  01:43:58.77  &  $-$71:50:09.7   &  18 	 &  8.89656	     &   0.06		   &	 12.350 		    &	      0      &  75\% CC few points at min  \\  
 002   &  01:43:15.53  &  $-$71:42:49.8   &  18 	 &  8.897351	     &   0.08		   &	 12.599 		    &	      0      &  75\% CC few points at min  \\  
 003   &  01:28:07.58  &  $-$72:48:52.1   &  18 	 &  12.52367	     &   0.27		   &	 12.758 		    &	      1      &  100\% CC GCVS2347 	  \\  
 004   &  01:24:25.35  &  $-$74:16:50.5   &  16 	 &  13.151942	     &   0.29		   &	 12.792 		    &	      1      &  100\% CC GCVS2343 	  \\  
 005   &  01:42:56.52  &  $-$71:18:46.0   &  18 	 &  9.122673	     &   0.05		   &	 12.885 		    &	      0      &  75\% CC few points at min  \\  
 006   &  01:38:18.83  &  $-$71:22:18.4   &  18 	 &  8.886835	     &   0.04		   &	 13.076 		    &	      0      &  75\% CC few points at min  \\  
 007   &  01:23:00.57  &  $-$74:22:16.8   &  16 	 &  9.752387	     &   0.21		   &	 13.182 		    &	      1      &  100\% CC GCVS2337 	  \\  
 008   &  01:38:00.64  &  $-$71:39:22.2   &  18 	 &  8.887862	     &   0.06		   &	 13.269 		    &	      0      &  75\% CC few points at min  \\  
 009   &  00:29:43.53  &  $-$71:33:21.0   &  17 	 &  8.371283	     &   0.06		   &	 13.470 		    &	      0      &  100\% CC 		  \\  
 010   &  01:13:25.40  &  $-$70:58:09.2   &  14 	 &  5.233301	     &   0.06		   &	 13.533 		    &	      0      &  75\% CC few points at min  \\  
\hline
\end{tabular}
\end{table*}

 During the revision phase of this manuscript, \cite{Sos15b} paper presenting  CCs in the Magellanic Clouds was posted as 
a preprint on the ArXiv. 
The authors comment that they can counter-identify 278 sources out of the 299 Cepheids in our list (the remaining 21 
fall in the inter-CCD gaps of their camera; Soszynski, private communication) and they confirm 35 
of the SMC Cepheids we have identified in our study. 
They also say that most of the remaining objects from our list turned out to be
constant or nearly constant in the optical bands.
We have cross-matched our  catalogue with \cite{Sos15b}'s and find that 
indeed 36 (about 13\%) of our Cepheids  are confirmed by OGLE~IV, namely, the nine CCs in common with the GCVS,  the two ACs 
also identified by  \cite{Sos15a} and 25 new CCs. 
This result is encouraging, since it confirms that our method
is promising despite the limited number of data points and the intrinsically low amplitude of the VMC
light curves. But it is also quite puzzling due to the low rate of optical confirmations, since 
our 299 bona fide Cepheids were selected not simply because of the light variation
in the $K_\mathrm{s}$-band, but, more importantly, because they also fall in the
colour-magnitude and colour-colour diagrams where OGLE III SMC
bona-fide Cepheids are found, and because they also follow the SMC 
Cepheid $PL$ relation. 

\begin{figure*}
\begin{centering}
\includegraphics[width=17cm,height=21cm]{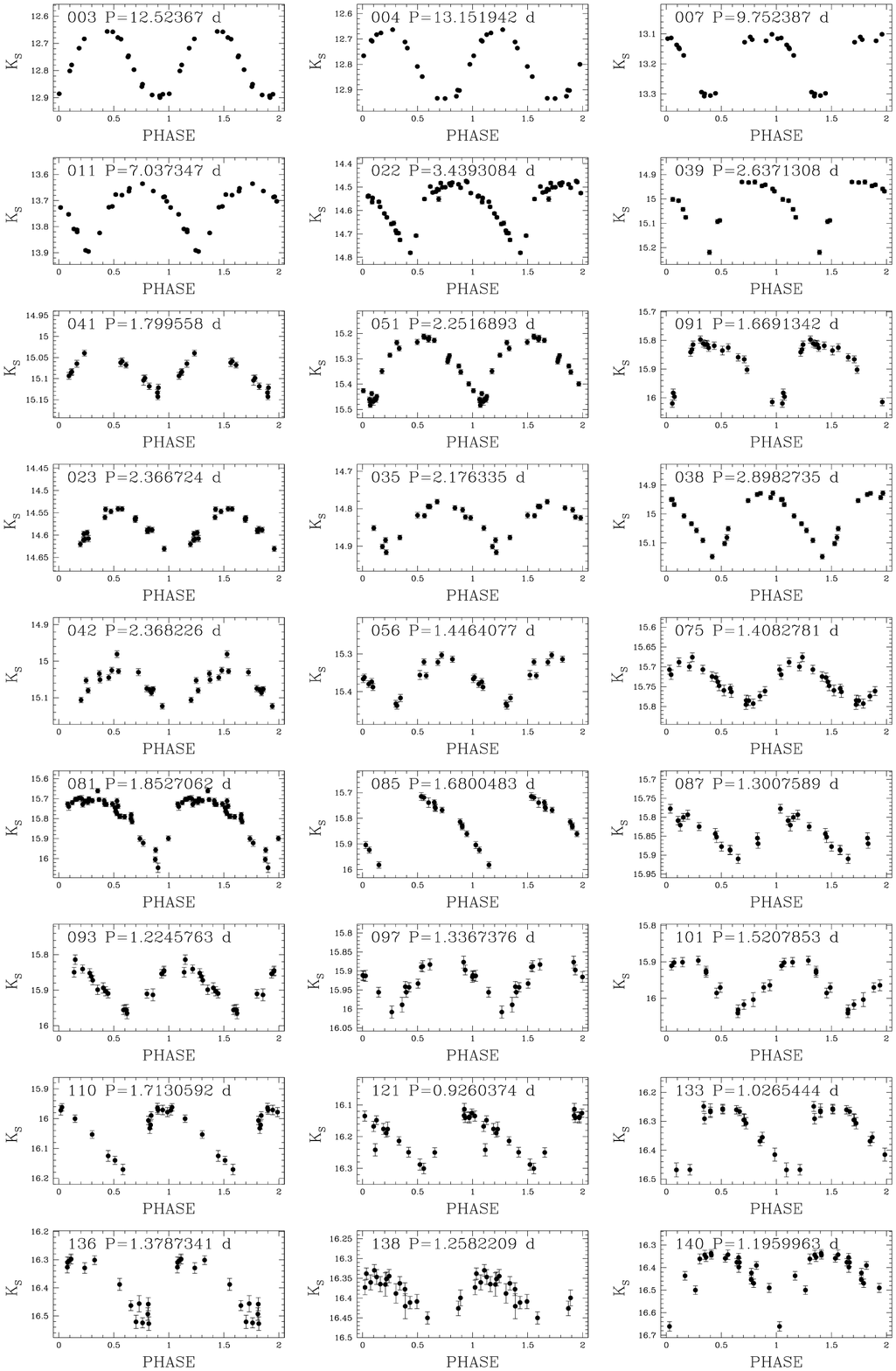} 
\caption{$K_\mathrm{s}$-band light curves of bona fide Cepheids  in the external regions of 
the SMC identified in the present study. 
For each source we indicate the VMC ID (ordered by increasing magnitude; see also Tab.~\ref{tab:CCcandidates1}) and 
the period derived from the analysis of the light curve with {\it SigSpec}. 
 We first display the 9 CCs in common with the GCVS (VMC-SMC-CEP-003, 004, 007, 011, 022, 039, 041, 051, 091) 
and 25 new CCs discovered in our study  that were later confirmed also by OGLE~IV  (VMC-SMC-CEP-023, 035, 038, 042, 056, 075, 081, 085, 
087, 093, 097, 101, 110, 121, 133, 136, 138, 140, 161, 
166, 177, 188, 231, 237, 292). All other bona fide Cepheids (including the two ACs) follow.
The figure is published in its entirety as Supporting Information with the electronic version of the article.}
\label{fig:lightCurves1}
\end{centering} 
\end{figure*}

The $K_\mathrm{s}$-band light curves of our 299 bona fide SMC Cepheids folded according to  the period derived 
from the analysis with {\it SigSpec}, are shown in Fig.~\ref{fig:lightCurves1}. 
We first display the 9 CCs in common with the GCVS 
and the 25 new CCs discovered in our study  and later confirmed also by OGLE~IV;  
all other bona fide Cepheids (including the two ACs) follow.
The light curves appear overall quite symmetrical and it is difficult with the VMC $K_\mathrm{s}$-band data to identify features such as bumps. 
$K_\mathrm{s}$ time series data for all of them 
 are provided in Table~\ref{tab:lightcurves}.
 Among the remarks  in the last column of Tab.~\ref{tab:CCcandidates1}, ``few points at min/max'' is used whenever the light 
curve is not homogeneously covered at minimum or maximum light. For some of these sources, we checked the light curve 
of nearby sources within 5 arcsec, to rule out unidentified photometric problems that might cause/mimic the light 
variation. They look flat suggesting that the variability of the sources identified as Cepheids is real, even for
 less well sampled light curves.

 Finally, we recall that according to the results in Section~\ref{sec:PL-selection} our technique enables recovery of 
about a 66\% of the true Cepheids that may occur in these external regions of the SMC and that  
 bona fide Cepheids may also be present in the sample of 
about 2,000 candidate variables fainter than $K_\mathrm{s}$ = 17.0 mag that were not analyzed here. 
Therefore,  there are likely additional 
Cepheids that we may have missed either 
because  their {\it SigSpec} periods are wrong by more than $\pm$ 2.5 d, thus 
causing them to fall  outside the 3$\sigma$ boundaries of 
the $PL$s\footnote{Indeed, there are about 30 CCs detected by OGLE~IV that are not in our list. 
 These sources were not identified because their SigSpec period is incorrect 
 (see Sections~\ref{sec:lc-analysis}, ~\ref{sec:PL-selection} for details), hence they did not pass the $PL$ selection.}, 
or because they are fainter than $K_\mathrm{s}$ = 17.0 mag.

\begin{table}
\begin{center}
\caption{$K_\mathrm{s}$-band time-series photometry of our 299 SMC Cepheids. 
The table is published in its entirety as Supporting 
Information with the electronic version of the article. A portion is
shown here for guidance regarding its form and content. 
}
\label{tab:lightcurves}
\begin{tabular}{ccc}
\hline
\hline
\multicolumn{3}{c}{Star VMC-SMC-CEP-001 } \\  
\hline 
HJD-2400000  &    $K_\mathrm{s}$     &    err$K_\mathrm{s}$      \\  
     (d)     &    (mag)              &    (mag)                 \\   
\hline
   55834.792559     &	 12.339     &     0.002      \\   
   55927.600518     &	 12.344     &     0.002      \\   
   56146.865651     &	 12.336     &     0.002      \\   
   56147.757493     &	 12.336     &     0.002      \\   
   56155.743101     &	 12.336     &     0.002      \\   
   56159.787169     &	 12.342     &     0.002      \\   
   56163.761878     &	 12.327     &     0.002      \\   
   56175.694234     &	 12.330     &     0.002      \\   
   56188.672502     &	 12.389     &     0.002      \\   
   56189.703199     &	 12.360     &     0.002      \\   
   56190.724434     &	 12.334     &     0.002      \\   
   56208.619677     &	 12.337     &     0.002      \\   
   56226.551935     &	 12.335     &     0.002      \\   
   56256.553723     &	 12.336     &     0.002      \\   
   56282.578397     &	 12.333     &     0.002      \\   
   56300.546758     &	 12.329     &     0.002      \\   
   56486.866850     &	 12.326     &	  0.002      \\  
   56911.795036     &	 12.331     &     0.002      \\  
\hline						   
\end{tabular}
\end{center}
\end{table}

\subsection{Contamination by other types of variable stars}\label{sec:contamination}
Adopting as reference the catalogue of  RR Lyrae stars  detected in the SMC by the OGLE~III 
survey  \citep{Sos10b},  Muraveva et al. (in prep.) 
studied the $K_\mathrm{s}$-band $PL$ relation of 1,081 SMC  RR Lyrae stars observed by VMC. They found  
that  these variables typically have  
 $K_\mathrm{s}$ mean magnitude between 17.5 mag and 19.5 mag, with a subset of about 200 of them 
having 17.5 $\leq  \langle K_\mathrm{s} \rangle \leq$ 18.2 mag.  
In this magnitude range there is partial overlap with the short period, faint end of the CCs distribution (see Fig.~\ref{fig:cmdKJK}). 
Furthermore, while the number  of CCs is expected to drop significantly moving from the centre to the external  region 
of the SMC, the distribution of RR Lyrae stars declines gently and their number is expected to remain rather high 
in the peripheral areas we are investigating (e.g. Fig. 7 of~\citealt{Sos15a}). Therefore, some of the 1,959 candidate variables  with $17.0 \leq \langle K_\mathrm{s} \rangle \leq $ 18.2 mag 
may be RR Lyrae stars. 

A further three percent contamination can also be expected from ECLs, they can spread 
all over the colour-magnitude diagram (see \citealt{Mor14}), 
thus we would need different criteria to distinguish them. 

Finally, some of the sources could be ACs like VMC-SMC-CEP-258 and 286,  or T2Ceps, due to the partial 
overlap existing among the $PL$s of the different types of Cepheids, especially between CCs and  
ACs (e.g. \citealt{Sos08b}). 

\section{Summary and conclusions}\label{sec:summary}
 We have developed a technique to identify variable stars using only the  multi-epoch 
near-infrared photometry obtained by the VMC survey and have specifically tailored it  to the identification of Cepheids. 
The technique exploits  colour-magnitude and colour-colour diagrams, and  $PL$ relations defined by known SMC Cepheids, along with template 
parameter selections and  visual inspection of the light curve to identify bona fide Cepheids. 
The technique was 
 applied  to external  regions of the SMC  
for which complete VMC $K_\mathrm{s}$-band observations  are  available and no comprehensive identification of variable 
stars from other surveys exists yet. 
 We have identified and present $K_\mathrm{s}$-band light curves for 299 SMC Cepheids,  of which 9 are  CCs in common with the General Catalog of Variable Stars (GCVS, \citealt{Art96}), two are ACs also found by  \cite{Sos15a} and the remaining 288 sources are new discoveries. The number of SMC Cepheids  we have detected is consistent with the predictions of \cite{Rub15} SFH recovery  in these regions of the SMC,  taking into account  that our technique may enable recovery of only 
about a 66\% of the true Cepheids that may occur in these external regions of the SMC.
Subsequently,   \cite{Sos15b} crossmatched  
278 of the sources in our list with their optical photometry and confirmed 36 of them as Cepheids.     
This result is encouraging, since it shows that our technique is promising despite the limited number of data points and the intrinsically low amplitude of the VMC
light curves, but  the low rate of optical confirmations is rather surprising and calls 
for further investigations to understand this discrepancy and possible physical/technical  reasons behind it.
This  near-infrared vs optical light-curve
connection/conspiracy will be addressed in a following paper.

\section*{Acknowledgments}
We thank the CASU and the WFAU for providing calibrated data products under the support of the
Science and Technology Facility Council (STFC) in the UK. Partial financial support for this
work was provided by PRIN-MIUR 2011 (PI: F. Matteucci).
 We thank the anonimous referee for his/her constructive comments.
RdG acknowledges funding support from the National Natural Science Foundation of China (grant 11373010).
 MIM thanks Felice Cusano and Alceste Bonanos for the interesting and useful discussions.
MRC acknowledges support from the UK's Science and Technology Facilities Council 
[grant number ST/M001008/1] and from the German Academic Exchange Service.

\end{document}